\documentclass[
    ,final            
  ]
  {aipproc}

\layoutstyle{6x9}
\usepackage{amssymb,amscd}
\usepackage[figuresright]{rotating}
\usepackage{epsfig}
\usepackage{array}
\usepackage{graphics}
\usepackage{graphicx}
\usepackage{multirow}
\usepackage{supertabular}
\usepackage{longtable}
\usepackage{dcolumn}
\usepackage{xspace}
\usepackage{tabularx}
\usepackage{color}
\usepackage{calc}
\usepackage{floatflt}

\begin{document}

\title{Impact of future HERA data on the determination of proton parton distribution functions using the ZEUS QCD fit}


\author{CLAIRE GWENLAN}{
  address={Nuclear \& Astrophysics Laboratory, University of Oxford, Keble Road, Oxford, OX1 3RH. UK.\\
           e-mail: c.gwenlan1@physics.ox.ac.uk}
}

\begin{abstract}
The high precision and large kinematic coverage of the data from the HERA-I running period (1994-2000) 
have already allowed precise extractions of proton parton distribution functions (PDFs). The 
HERA-II running program is now underway and is expected to 
provide a substantial increase in the luminosity collected at HERA. In this paper, a study is 
presented which investigates the potential impact of future data from HERA on the proton 
PDF uncertainties, within the currently planned running scenario. In addition, the effect 
of a possible future measurement of the longitudinal structure function, $F_{\rm L}$, on the gluon 
distribution is investigated.
\end{abstract}

\maketitle

\vspace{-1.2cm}
\section{Introduction}
\vspace{-0.25cm}
Since the advent of HERA, much progress has been made in determining the Parton Distribution Functions 
(PDFs) of the proton. The PDFs must 
be known as precisely as possible in order to make reliable predictions for any process involving protons, 
and to maximise the discovery potential for new physics at both current and future colliders.

HERA is now in its second stage of operation.
With the measurements that can now be expected from HERA-II, our knowledge of PDFs 
should be further improved. In this paper, first studies of the potential impact of 
future measurements from HERA, on the PDF uncertainties, are presented. 

\vspace{-0.2cm}
\section{HERA Physics and Kinematics}
\vspace{-0.25cm}
Lepton-proton Deep Inelastic Scattering (DIS) can proceed either via the Neutral Current (NC) 
interaction (through the exchange of a $\gamma^*$ or ${\rm Z}^0$), or via the Charged Current 
(CC) interaction (through the exchange of a ${\rm W}^{\pm}$). The kinematics of lepton-proton DIS are described 
in terms of the Bjorken scaling variable, $x$, the negative invariant mass squared of the 
exchanged vector boson, $Q^2$, and the fraction of energy transferred from the lepton to the 
hadron system, $y$. 
The three quantities are related by $Q^2=sxy$, where $s$ is the centre-of-mass energy squared.

At leading order (LO) in the electroweak interaction, the double 
differential cross section for the NC DIS process is given in terms of proton structure functions,
\begin{eqnarray}
\frac{{\rm d}^2\sigma^{\rm NC}(e^{\pm}p)}{{\rm d}x{\rm d}Q^2}\sim Y_+ {\rm F}_{2}(x,Q^2) - y^2{\rm F}_{\rm L}(x,Q^2) \mp Y_- x{\rm F}_{3}(x,Q^2)  
\label{Eqn:NC}
\vspace{-0.2cm}
\end{eqnarray}
where $Y_{\pm}=1\pm(1-y)^2$. 
The structure functions are directly related to the PDFs and their $Q^2$ dependence 
is predicted in perturbative QCD. 
In particular, ${\rm F}_2$ and $x{\rm F}_3$ depend directly on the quark distributions. 
For $x < 10^{-2}$, ${\rm F}_2$ is dominated by sea quarks and the $Q^2$ dependence is 
driven by gluon radiation. At high $Q^2 \gtrsim M_{\rm Z}^2$, the contribution from $x{\rm F}_3$ 
is significant. The longitudinal structure function, 
${\rm F}_{\rm L}$, is directly sensitive to the gluon, 
but is only important at high-$y$. At LO, the CC cross sections are given by,
\begin{eqnarray}
~~~~\frac{{\rm d}^2\sigma^{\rm CC}(e^{+}p)}{{\rm d}x{\rm d}Q^2} \sim \bar{u}+\bar{c}+(1-y)^2(d+s);~~~~~~ \frac{{\rm d}^2\sigma^{\rm CC}(e^{-}p)}{{\rm d}x{\rm d}Q^2} \sim {u}+{c}+(1-y)^2(\bar{d}+\bar{s})\nonumber 
\vspace{-0.2cm}
\end{eqnarray}
so that a measurement of the $e^+p$ and $e^-p$ cross sections at high-$x$, provide information on 
the $d$- and $u$-valence quarks, respectively, thereby allowing the separation of flavour. 
\begin{table}
\begin{tabular}{llcc}
\hline
    \tablehead{1}{c}{b}{data sample\\}
  &  \tablehead{1}{c}{b}{kinematic coverage\\}
  & \tablehead{1}{c}{b}{HERA-I \\$\mathcal{L}$ (${\rm pb^{-1}}$)\\}
  & \tablehead{1}{c}{b}{HERA-II \\$\mathcal{L}$ (${\rm pb^{-1}}$)\\(assumed)}\\
\hline
 96-97 NC $e^+p$~\cite{epj:c21:443} &  $2.7<Q^2<30000$ ${\rm GeV}^2$; $6.3 \cdot 10^{-5}<x<0.65$    &30  &30  \\
 94-97 CC $e^+p$~\cite{epj:c12:411} &  $280<Q^2<17000$ ${\rm GeV}^2$; $6.3 \cdot 10^{-5}<x<0.65$  & 48 &48\\
 98-99 NC $e^-p$~\cite{epj:c28:175} &  $200<Q^2<30000$ ${\rm GeV}^2$; $0.005<x<0.65$  &  16 &350 \\
 98-99 CC $e^-p$~\cite{pl:b539:197} &  $280<Q^2<17000$ ${\rm GeV}^2$; $0.015<x<0.42$  &  16 &350 \\
 99-00 NC $e^+p$~\cite{hep-ex:0401003} & $200<Q^2<30000$ ${\rm GeV}^2$; $0.005<x<0.65$ &  63 &350\\
 99-00 CC $e^+p$~\cite{epj:c32:16} &  $280<Q^2<17000$ ${\rm GeV}^2$; $0.008<x<0.42$    &  61 &350 \\
 96-97 inc. DIS jets~\cite{pl:b547:164}  &  $125<Q^2<30000$ ${\rm GeV}^2$; $E_{\rm T}^{Breit}>8$ GeV   &  37  &500 \\
 96-97 dijets in $\gamma p$~\cite{epj:c23:615} &   $Q^2 \lesssim 1$ ${\rm GeV}^2$; $E_{\rm T}^{jet1,2}>14,11$ ${\rm GeV}$   &  37  &500 \\ \hline
 optimised jets~\cite{chris} &   $Q^2 \lesssim 1$ ${\rm GeV}^2$; $E_{\rm T}^{jet1,2}>20,15$ ${\rm GeV}$    &  -  &500 \\
\hline
\end{tabular}
\caption{The data-sets included in the ZEUS PDF fits. 
The last two columns give the integrated luminosities of the HERA-I measurements and those assumed 
in the HERA-II projection. Note that the 96-97 NC and the 94-97 CC measurements have not had their luminosity scaled. 
}
\label{tab:HERAIIASSUMPTIONS}
\end{table}

The QCD scaling violations in the inclusive cross section data, namely the QCD 
Compton ($\gamma^{*}q \rightarrow gq$) and boson-gluon-fusion 
($\gamma^{*}g \rightarrow q\bar{q}$) processes, may also give rise to distinct 
jets in the final state.
Jet cross sections therefore provide a direct 
constraint on the gluon through the boson-gluon-fusion process.
\begin{figure}[Ht]
\includegraphics[width=15cm,height=9cm]{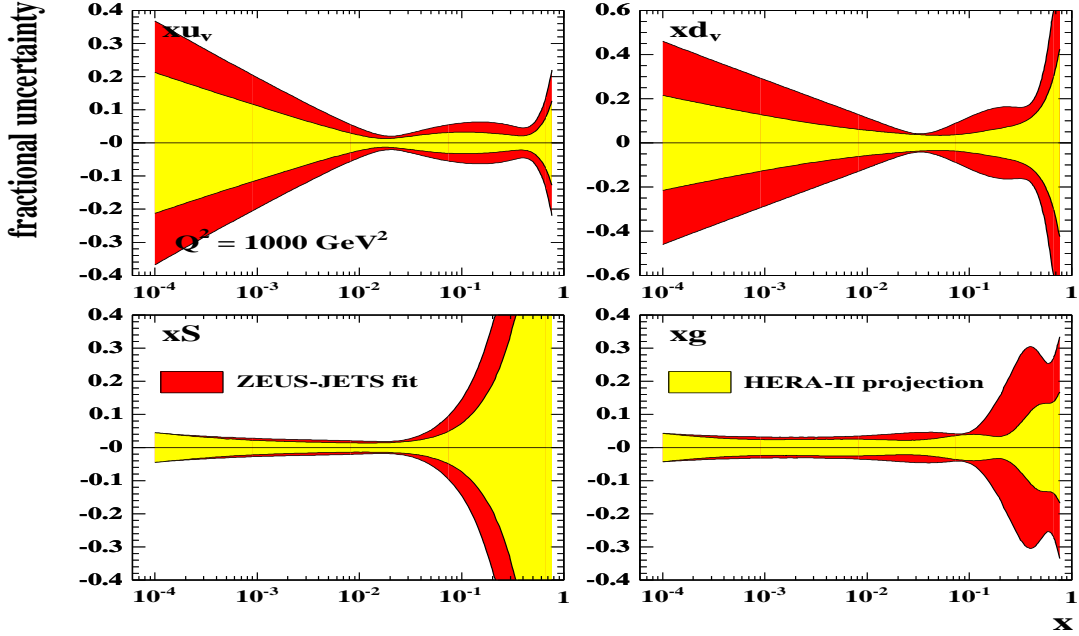}
\caption{The fractional PDF uncertainties, as a function of $x$, for the $u$-valence, 
$d$-valence, sea-quark and gluon distributions at $Q^2=$ 1000 ${\rm GeV}^2$. 
The red shaded bands show the results of the ZEUS-JETS fit and the yellow shaded bands show the 
results of the HERA-II projected fit.}
\label{Fig:PDFS}
\end{figure}

\vspace{-0.2cm}
\section{Results and Discussion}
\vspace{-0.25cm}
In this section, the results of two separate studies are presented. The first study, 
provides an estimate of how well the PDF uncetrainties may be known by the end of HERA-II, 
within the currently planned running scenario, while the second study investigates the impact of a future HERA 
measurement of ${\rm F_L}$ on the gluon uncertainty.

All results presented, are based on the recent ZEUS-JETS PDF analysis~\cite{epj:c050364}, 
which is performed in the conventional next-to-leading-order (NLO) DGLAP~\cite{dglap} framework of QCD.
The fit includes the full set of ZEUS inclusive NC~\cite{epj:c21:443,epj:c28:175,hep-ex:0401003} 
and CC~\cite{epj:c12:411,pl:b539:197,epj:c32:16} data from HERA-I, 
as well as ZEUS inclusive jet data in DIS~\cite{pl:b547:164} and dijets in $\gamma$p collisions~\cite{epj:c23:615}. 
The ZEUS-JETS fit uses the offset method~\cite{hep-ph:0205153} for the evaluation of PDF uncertainties.

{{\bf 1. PDF uncertainty estimates for the end of HERA running}}
\vspace{0.05cm}

The data from HERA-I (1994-2000) are already very precise and cover a wide kinematic region.
However, HERA-II is now running efficiently and is expected to provide a substantial increase in luminosity. 
Current estimates suggest that, by the end of HERA running, an integrated luminosity of $700$ ${\rm pb}^{-1}$ should be achievable. 
This will allow more precise measurements of cross sections 
that are curently statistically limited: in particular, the 
high-$Q^2$ NC and CC data, as well as high-$Q^2$ and/or high-$E_{\rm T}$ jet data. In addition to the simple increase in luminosity, 
recent studies~\cite{chris} have shown that future jet cross section measurements, in kinematic regions optimised for sensitivity to PDFs, should 
have a significant impact on the gluon uncertainties.

In this contribution, the effect on the PDF uncertainties, 
of both the higher precision expected from HERA-II and the possibility of optimised jet cross section measurements,
has been estimated in a new QCD fit. This fit will be referred to as the 'HERA-II projection', 
In the HERA-II projected fit, the statistical uncertainties on the currently available HERA-I data have been reduced.
For the high-$Q^2$ inclusive data, a total integrated luminosity of $700$ ${\rm pb}^{-1}$ was assumed, equally divided between $e^+$ and $e^-$. 
For the jet data, an integrated luminosity of $500$ ${\rm pb}^{-1}$ was assumed. The central values and systematic uncertainties were 
taken from the published data in each case. In addition to the assumed increase in precision of the measurements, 
a set of optimised jet cross sections were also included, for forward dijets in $\gamma p$ collisions, as defined in a recent study~\cite{chris}. Since no real 
data are yet available, simulated points were calculated using the NLO QCD program of Frixione-Ridolfi~\cite{np:b507}, 
with statistical uncertainties corresponding to $500$ ${\rm pb}^{-1}$. For this study, systematic uncertainties on 
the optimised jet cross sections were ignored. Table~\ref{tab:HERAIIASSUMPTIONS} summarises the data-sets included in 
the fit, and gives the luminosities of the (real) HERA-I measurements compared to those assumed in the HERA-II projection.

The results are summarised in Fig.~\ref{Fig:PDFS}, which shows the fractional PDF uncertainties, for the $u$- and $d$-valences, 
sea-quark and gluon distributions, at $Q^2 = 1000$ ${\rm GeV}^2$, for the ZEUS-JETS fit compared to the HERA-II projection. 
Note that the same general features are observed for all values of $Q^2$. 
In fits to only HERA data, the information on the valence quarks comes from the high-$Q^2$ NC and CC cross sections. 
The increased statistical precision of the high-$Q^2$ data, as assumed in the HERA-II projected fit, 
gives a significant improvement in the valence uncertainties over the whole range of $x$. For the sea 
quarks, a significant improvement in the uncertainties at high-$x$ is also observed. In contrast, the low-$x$ 
uncertainties are not visibly reduced. This is due to the fact that the data constraining the low-$x$ region 
tends to be at lower-$Q^2$, which are already systematically limited. This is also the reason why the low-$x$ gluon 
uncertainties are not significantly reduced. However, the mid-to-high-$x$ gluon, which is constrained 
by the jet data, is much improved in the HERA-II projected fit. Note that about half 
of the observed reduction in the gluon uncertainties is due to the inclusion of the simulated optimised jet cross sections.
The improvement to the high-$x$ partons observed in the HERA-II projected fit, 
is particularly relevant for high-scale physics at the LHC e.g. high-$E_{\rm T}$ jets, new particle searches.

\vspace{0.4cm}
{\bf 2. Impact of a future HERA measurement of $F_{\rm L}$ on the gluon PDF}
\vspace{0.05cm}

The longitudinal structure function, ${\rm F_L}$, is directly related to the gluon density in 
the proton.
In principle, ${\rm F}_{\rm L}$ can be extracted by measuring the NC DIS cross section at fixed $x$ and $Q^2$, 
for different values of $y$ (see Eqn.~\ref{Eqn:NC}). A precision measurement could be achieved by varying the centre-of-mass 
energy, since $s=Q^2/xy\approx4E_eE_p$, where $E_e$ and $E_p$ are the electron and proton beam energies, 
respectively. Studies~\cite{max:fl} have shown that this would be most efficiently achieved by 
changing the proton beam energy. However, such a measurement has not yet been performed at HERA.

For the present study, the impact of a possible future HERA measurement of ${\rm F_L}$ on the gluon PDF uncertainties 
has been investigated, using a set of simulated ${\rm F_L}$ data-points~\cite{max:fl}. 
The simulation was performed using the GRV94~\cite{zphys:c67:433} proton PDF for the central values, and 
assuming $E_e = 27.6$ GeV and $E_p = 920, 575, 465$ and $400$ GeV, with luminosities of 10, 5, 3 and 2 ${\rm pb}^{-1}$, respectively. 
Under such conditions, and assuming the luminosity scales as $E_p^2$, this scenario would nominally 
cost $35$ ${\rm pb}^{-1}$ of luminosity under standard HERA conditions. However, this takes no account of time taken 
for optimisation of the machine with each change in $E_p$, which could be considerable. The systematic uncertainties 
on the simulated data-points were calculated assuming a $\sim 2\%$ precision on the inclusive NC 
cross section measurement. A more comprehensive description of the simulated data is given elsewhere~\cite{max:fl}.
\begin{figure}[Ht]
\includegraphics[width=15cm,height=9.cm]{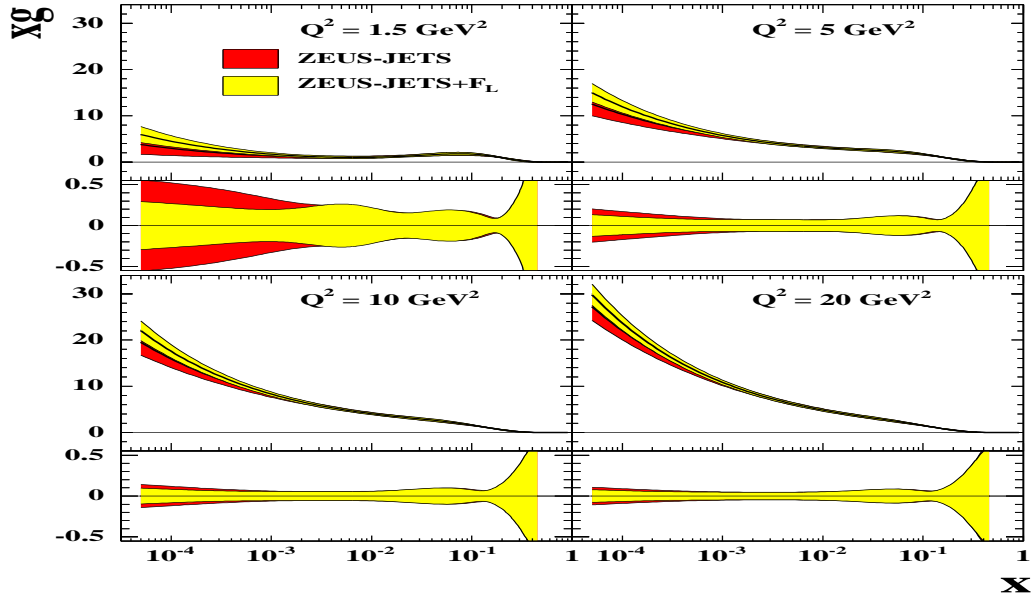}
\caption{The gluon PDFs, showing also the fractional uncertainty, for fits with and 
without inclusion of the simulated ${\rm F_L}$ data, for $Q^2 =$ 1.5, 5, 10 
and 20 ${\rm GeV}^2$. The red shaded bands show the results of the ZEUS-JETS fit and the 
yellow shadeds band show the results of the ZEUS-JETS+${\rm F_L}$ fit.}
\label{Fig:FL}
\end{figure}

The simulated data were included in the ZEUS-JETS fit. 
Figure~\ref{Fig:FL} shows the gluon distribution and fractional uncertainties for fits with and without 
inclusion of the simulated ${\rm F_L}$ data. The 
results indicate that the gluon uncertainties are reduced at low-$x$, but the improvement is only significant at relatively low $Q^2$. 

\vspace{-0.15cm}
\section{Acknowledgements}
\vspace{-0.2cm}
I thank C. Targett-Adams for providing the NLO QCD jet cross section predictions and the machinery 
to allow their inclusion in the ZEUS fit. I also thank 
M. Klein and R. Thorne for providing ${\rm F_L}$ predictions. This work was supported by PPARC.

\end{document}